\documentclass[a4paper]{article}

\usepackage{JCMS}           % load package
\usepackage[utf8]{inputenc} % allow utf-8 input
\usepackage[T1]{fontenc}    % use 8-bit T1 fonts
\usepackage{type1cm}        %
\usepackage{booktabs}       % professional quality tables
\usepackage{amsfonts}       % blackboard math symbols
\usepackage{nicefrac}       % compact symbols for 1/2, etc.
\usepackage{microtype}      % microtypography
\usepackage{graphicx}       % figures
\usepackage{csquotes}       % required for biblatex-apa
\usepackage[backend=biber,style=apa]{biblatex} % for required APA7 referencing
\DefineBibliographyStrings{english}{
  andothers = {\mkbibemph{et\addabbrvspace al\adddot}}
}
\usepackage{url}            % simple URL typesetting
\usepackage{hyperref}       % hyperlinks

\usepackage{multirow}
\usepackage{amsmath}
\usepackage{musicography}

\title{BERT-like Pre-training for Symbolic Piano Music Classification Tasks}

\author{%
  Yi-Hui Chou\thanks{The first two authors contribute equally to the paper.} \\
  Carnegie Mellon University, United States \\
  \texttt{yihuic@andrew.cmu.edu} \\
  \And
  I-Chun Chen$^{\ast}$ \\
  National Tsing Hua University, Taiwan \\
  \texttt{icchen0101@elsa.cs.nthu.edu.tw} \\
  \AND
  Chin-Jui Chang \\
  Research Center for IT Innovation, Academia Sinica, Taiwan \\
  \texttt{csc63182@citi.sinica.edu.tw} \\
  \And
  Joann Ching \\
  Research Center for IT Innovation, Academia Sinica, Taiwan \\
  \texttt{joann8512@citi.sinica.edu.tw} \\
  \And
  Yi-Hsuan Yang \\
  National Taiwan University, Taiwan \\
  \texttt{yhyangtw@ntu.edu.tw} \\
}

\begin{document}

\maketitle

\begin{abstract}
This article presents a benchmark study of symbolic piano music classification using the masked language modelling approach of the Bidirectional Encoder Representations from Transformers (BERT). Specifically, we consider two types of MIDI data:  \emph{MIDI scores}, which are musical scores rendered directly into MIDI with no dynamics and precisely aligned with the metrical grid notated by its composer and  \emph{MIDI performances}, which are MIDI encodings of human performances of musical scoresheets. 
With five public-domain datasets of single-track piano MIDI files, we pre-train two 12-layer Transformer models using the BERT approach, one for MIDI scores and the other for MIDI performances, and fine-tune them for four downstream classification tasks. These include two note-level classification tasks (melody extraction and velocity prediction) and two sequence-level classification tasks (style classification and emotion classification). 
Our evaluation shows that the BERT approach leads to higher classification accuracy than recurrent neural network (RNN)-based baselines. 
\newline

\end{abstract}

\section{Introduction}

In the literature of machine learning, a prominent approach to overcome the labelled data scarcity issue is to adopt ``transfer learning'' and divide the learning problem into two stages \parencite{han2021pretrained}: a \emph{pre-training} stage that establishes a model capturing general knowledge from one or multiple source tasks and a \emph{fine-tuning} stage that transfers the captured knowledge to target tasks. 
Model pre-training can be done in two ways: using a labelled dataset \parencite{choi2017transfer,kim19,9414405}, such as training a VGG-like model over millions of human-labelled clips of general sound events and then fine-tuning it on instrument classification  \parencite{gururani19ismir}; or using an unlabelled dataset with a self-supervised training strategy.
%Model pre-training can be done using either a labelled dataset \parencite{choi2017transfer,kim19,9414405}, e.g., to train a VGG-like model over millions of human-labelled clips of general sound events, and then fine-tune it on instrument classification data \parencite{gururani19ismir}. Alternatively, it can utilize an unlabelled dataset with a self-supervised training strategy. 
The latter is in particular popular in the field of natural language processing (NLP), where pre-trained models (PTMs) using Transformers \parencite{vaswani2017attention} have achieved state-of-the-art results on almost all NLP tasks, including generative and discriminative ones \parencite{han2021pretrained}.

This article presents an empirical study employing PTMs to symbolic-domain piano music classification tasks. 
In particular,  inspired by the growing trend of treating MIDI music as a ``language'' in deep generative models for symbolic music \parencite{huang2018music,payne2019musenet,huang2020pop,musemorphose,musecoco},  
we employ a Transformer-based network pre-trained by a self-supervised training strategy called ``masked language modelling'' (MLM), which has been widely used in BERT-like PTMs in NLP \parencite{bert,roberta,spanbert,xlnet,cross-lingual-lm}. 
Despite the fame of BERT,  we are aware of only two publications that employ BERT-like PTMs for symbolic music classification \parencite{tsai20ismir,musicbert}. 
The first work \parencite{tsai20ismir} deals with optically scanned sheet music, while we use MIDI inputs.  
The second work \parencite{musicbert} uses a diverse set of multi-track MIDI files, while we intend to focus on piano music only.
We discuss how our work differs from these two existing works in more detail in Section \ref{sec:related_work}.

We evaluate PTMs on four piano music classification tasks. 
These include two \emph{note-level classification tasks}, i.e., melody extraction \parencite{simonettaCNW19,note-affinity} and velocity prediction \parencite{widmer94aaai,jeongKKLN19ismir,jeongKKN19icml} and two \emph{sequence-level classification tasks}, i.e., style classification \parencite{lee20ismirLBD,kong2020largescale} and emotion classification \parencite{grekow2009detecting,lin2013exploration,panda2013multi,panda2018}. 
We use five datasets in this work, amounting to 4,166 pieces of piano MIDI.
We give details of the datasets and tasks in Sections \ref{sec:db} and  \ref{sec:task}.

As the major contribution of this article, we report a performance study of variants of PTM for this diverse set of classification tasks, comparing the proposed approach (Section \ref{sec:bert}) with recurrent neural network (RNN)-based baselines (Section \ref{sec:baseline}).
Results reported in Section \ref{sec:exp}  show that the ``pre-train and fine-tune'' strategy does lead to higher accuracy than the RNN baselines.
Moreover, we consider two types of MIDI data and compare the performance of the resulting PTMs. Specifically, following \textcite{oore2018time}, we differentiate two types of MIDI files, \emph{MIDI scores}, which are musical scoresheets rendered directly into MIDI with no dynamics and exactly according to the written metrical grid, and \emph{MIDI performances}, which are MIDI encodings of human performances of musical scoresheets.  
All the 4,166 pieces we have are MIDI performances, but we can obtain the corresponding MIDI-score version of them by dropping performance-related information. Accordingly, we build two PTMs, one for MIDI scores and the other for MIDI performances and evaluate their performance respectively on the downstream tasks. While the MIDI-score version can be applied to a wider array of tasks involving those with or without performance-related information, the MIDI-performance version can likely perform better for tasks that involve human performance of piano scores, such as style classification and emotion classification. 
Therefore, such a performance comparison is relevant.

As the secondary contribution, we share the code and release checkpoints of the pre-trained and fine-tuned models publicly in our GitHub repository\footnote{\url{https://github.com/wazenmai/MIDI-BERT}} with an open-source licence.
Together with the fact that all the datasets employed in this work are publicly available, our research can be taken as a new testbed of PTMs in general and a new public benchmark for machine learning-based classification of MIDI music.

\section{Related Work on Pre-trained Models for MIDI}
\label{sec:related_work}

Machine learning has been applied to music in symbolic formats such as MIDI. Exemplary tasks include symbolic-domain music genre classification \parencite{correa16survey,ferraro18}, composer classification \parencite{lee20ismirLBD,kong2020largescale},
and melody note identification \parencite{simonettaCNW19, note-affinity}.
However, labelled datasets for symbolic-domain music data tend to be small in size in general \parencite{gttm14ismir,simonettaCNW19,harasim20ismir}, posing challenges to train effective supervised machine learning models that generalise well.

To our best knowledge, the work of \textcite{tsai20ismir} represents the first attempt to use PTMs for symbolic-domain music classification. They showed that either a RoBERTa-based Transformer encoder PTM \parencite{roberta} or a GPT2-based Transformer encoder PTM \parencite{gpt2} outperform non-pre-trained baselines for a 9-class symbolic-domain composer classification task. Pre-training boosts the classification accuracy for the GPT2 model greatly from 46\% to 70\%. However, the symbolic data format considered in their work is ``sheet music image'' \parencite{tsai20ismir}, which are images of musical scores. This data format has been much less used than MIDI in the literature. 

\textcite{musicbert} presented MusicBERT, a PTM tailored for symbolic MIDI data. MusicBERT was trained on a non-public dataset of over one million multi-track MIDI pieces. The authors showcased the efficacy of MusicBERT by applying it to two generative music tasks, melody completion and accompaniment suggestion and two sequence-level discriminative tasks, including genre and style classification. In comparison to non PTM-based baselines, MusicBERT consistently led to better performance.  
Our work differs from theirs in the following aspects. 
First, our pre-training corpus is much smaller (only 4,166 pieces) but all publicly available, less diverse but more dedicated (to piano music).
Second, we aim at establishing a benchmark for symbolic music classification and include not only sequence-level but also note-level tasks. Furthermore, the labelled data we employ for our downstream tasks is comparatively modest, with each dataset containing fewer than 1,000 annotated pieces.  This differs from MusicBERT's dataset, referred to as the TOP-MAGD dataset \parencite{ferraro18}, which comprises over 20,000 annotated pieces---a considerably extensive collection rarely encountered in symbolic music tasks.
Finally, their token representation is designed for multi-track MIDI, while ours is for single-track piano MIDI, each MIDI file is an individual movement of a longer work.

\begin{table*}[t]
\caption{Public datasets used for this article. All the datasets are used for pre-training, while three are also used for downstream classification tasks. Average note pitch is in MIDI number. The symbol ``\#'' stands for ``number of''.\newline}
\label{tab:datasets}
\centering
\resizebox{\textwidth}{!}{%
    \begin{tabular}{l l r r c c c c}
    \toprule
    \multirow{2}{*}{Dataset} & Downstream Classifi- & \multirow{2}{*}{Pieces} & Duration & Avg. note  & Avg. note  & Avg. \#notes & Avg. \#bars  \\
    & cation (CLS) Tasks&&(hours)&pitch& duration (in \musThirtySecond)&per bar& per piece \\
    \midrule
    Pop1K7 & - & 1,747& 108.8  & E4~~ & ~~8.5  & 16.9 & 103.3 \\ 
    ASAP$_\text{4/4}$ & - & 65 & 3.5  & D4\# & ~~2.9 & 23.0 & ~95.9 \\
    POP909$_\text{4/4}$ & melody, velocity  & 865 &  59.7  & D4\# & ~~6.1 & 17.4  & ~94.9 \\ 
    Pianist8 & style  & 411 & 31.9  & D4\# & ~~9.6 & 17.0 & 108.9 \\
    EMOPIA & emotion  & 1,078 & 12.0  & C4\# & 10.0 & 17.9 & ~14.8 \\ 
    \bottomrule
    \end{tabular}%
}
\end{table*}

\section{Datasets and Data Representation}
\label{sec:db}

\subsection{Piano MIDI Datasets}
\label{sec:db2}

We collect four existing public-domain piano MIDI datasets, including Pop1K7, ASAP, POP909, EMOPIA and compile ourselves a new dataset, named Pianist8.
To simplify the token representation, we consider only 
MIDI files that specify 4/4 metre.\footnote{We note that the metre can be wrong due to errors in automatic music transcription, leading to noise in the data. Future work can be done to improve this. Moreover, future work can be done to use a more complicated token representation such as that proposed by \textcite{ashis19ismir} to include other time signatures.}
We list some important statistics of these five datasets in Tab.~\ref{tab:datasets} and provide their details below.
\begin{itemize}
    \item The \textbf{Pop1K7} dataset developed by \textcite{hsiao21aaai}\footnote{\url{https://github.com/YatingMusic/compound-word-transformer}} is composed of machine transcriptions of 1,747 audio recordings of piano covers (i.e., a new recording by someone other than the original artist or composer of a commercially released song) of Japanese anime, Korean and Western pop music, amounting to over 100 hours worth of data.
    The transcription was done with the ``onsets-and-frames'' RNN-based piano transcription model \parencite{hawthorne2018onsets} 
    (which is reported to attain 95.32 onset F1 score for note-level piano transcription \parencite{hawthorne21ismir}),
    and the RNN-based downbeat and beat tracking model from the Madmom library \parencite{madmom}.
    This dataset is the largest among the five, constituting half of our training data. 
    We only use it for pre-training.
    
    \item \textbf{ASAP}, the aligned scores \& performances dataset compiled by \citeyear{asap},\footnote{\url{https://github.com/fosfrancesco/asap-dataset}} contains 1,068 MIDI performances of 222 Western classical music compositions from 15 composers, along with the MIDI performances of the 222 pieces compiled from the MAESTRO dataset \parencite{hawthorne19iclr}.
    We consider it as an additional dataset for pre-training, using only the MIDI that specifies 4/4 metre with no metre change at all throughout the piece.
    This leaves us with 65 pieces of MIDI files, which last for 3.5 hours in total.
    Tab. \ref{tab:datasets} shows that, being the only classical dataset among the five, ASAP features shorter average note duration and larger number of notes per bar.
    
    \item \textbf{POP909} comprises piano covers of 909 pop songs compiled by \textcite{pop909}.\footnote{\url{https://github.com/music-x-lab/POP909-Dataset}} It is the only dataset among the five that provides melody, non-melody labels for each note. Specifically, each note is labelled with one of the following three classes: \texttt{vocal melody} (piano notes corresponding to the lead vocal melody line in the original pop song, usually during the verse and chorus parts); \texttt{instrumental melody} (piano notes corresponding to the secondary melody line played by the instruments in the original pop song, usually during the intro, bridge, outro parts); and \texttt{accompaniment} (including arpeggios, broken chords and many other textures).\footnote{POP909 originally refers to \texttt{vocal melody} as \texttt{melody} and \texttt{instrumental melody} as \texttt{bridge} in their work \parencite{pop909}. We opt for our new naming to highlight the fact that the latter is also a type of melody.}
    As it is a MIDI performance dataset, it also comes with velocity information. Therefore, we use it for the melody classification and velocity prediction tasks. 
    We discard pieces that do not specify 4/4 metre, ending up with 865 pieces for this dataset.
    
    \item \textbf{Pianist8} consists of eight artists' performances of piano music that we download from YouTube for training and evaluating symbolic-domain style classification.\footnote{\url{https://zenodo.org/record/5089279}}  
    The artists are Richard Clayderman (pop), Yiruma (pop), Herbie Hancock (jazz), Ludovico Einaudi (contemporary), Hisaishi Joe (contemporary), Ryuichi Sakamoto (contemporary),  Bethel Music (religious) and Hillsong Worship (religious).
    The artists are also the composers of the pieces, except for Richard Clayderman, Bethel Music and Hillsong Worship.
    The dataset contains a total of 411 pieces, with a fairly balanced number of pieces per artist.
    Each audio file is paired with its MIDI performance, which is machine-transcribed by the piano transcription model proposed by \textcite{TTtranscription}.
    
    \item \textbf{EMOPIA} is a dataset of pop piano music collected recently by \textcite{emopia}  from YouTube for research on emotion-related tasks.\footnote{\url{https://annahung31.github.io/EMOPIA/}}  
    It has 1,087 clips (each around 30 seconds) segmented from 387 songs, covering Japanese anime, Korean \& Western pop song covers, movie soundtracks and personal compositions.
    The emotion of each clip has been labelled using the following 4-class taxonomy: \texttt{HAHV} (high arousal high valence); \texttt{LAHV} (low arousal high valence); \texttt{HALV} (high arousal low valence); and \texttt{LALV} (low arousal low valence). This taxonomy is derived from the Russell's valence-arousal model of emotion \parencite{russell}, where \emph{valence} indicates whether the emotion is positive or negative and \emph{arousal} denotes whether the emotion is high (e.g., angry) or low (e.g., sad) \parencite{yang11book}.
    The MIDI performances of these clips are similarly machine-transcribed from the audio recordings by the model of \textcite{TTtranscription}. 
    We use this dataset for the emotion classification task. As Tab. \ref{tab:datasets} shows, the average length of the pieces in the EMOPIA dataset is the shortest among the five, since they are actually clips manually selected by dedicated annotators \parencite{emopia} to ensure that each performance expresses a single emotion.
\end{itemize}

All five datasets consist of MIDI performances. As mentioned in the introduction, we intend to build two PTMs, one for MIDI scores and the other for MIDI performances. 
We obtain the MIDI-score version of each performance
by dropping velocity and tempo information, temporally quantising the onset time and duration of each the notes to the semiquaver resolution.

\subsection{Token Representation}
\label{sec:token}

Similar to text, a piece of music in MIDI can be considered as a sequence of musical events or ``tokens''. However, what makes music different is that musical notes are associated with a temporal length (i.e., note duration) and multiple notes can be played at the same time. Therefore, to represent music, we need \emph{note-related} tokens describing, for example, the pitch and duration of the notes, as well as \emph{metric-related} tokens placing the notes over a time grid.

In the literature, a variety of token representations for MIDI have been proposed, differing in many aspects such as the MIDI data being considered (e.g., melody \parencite{Magenta}, lead sheet \parencite{jazzTransformer20ismir}, piano \parencite{huang2018music} and multi-track music \parencite{payne2019musenet,multitrackmusictransformer23icassp}), the temporal resolution of the time grid and the way the advancement in time is notated \parencite{huang2020pop}. Auxiliary tokens describing, for example, the chord progression \parencite{huang2020pop} or grooving pattern \parencite{guitarTransformer20ismir} underlying a piece can also be added.

\begin{figure}[t]
\centering
\includegraphics[width=1.0\textwidth]{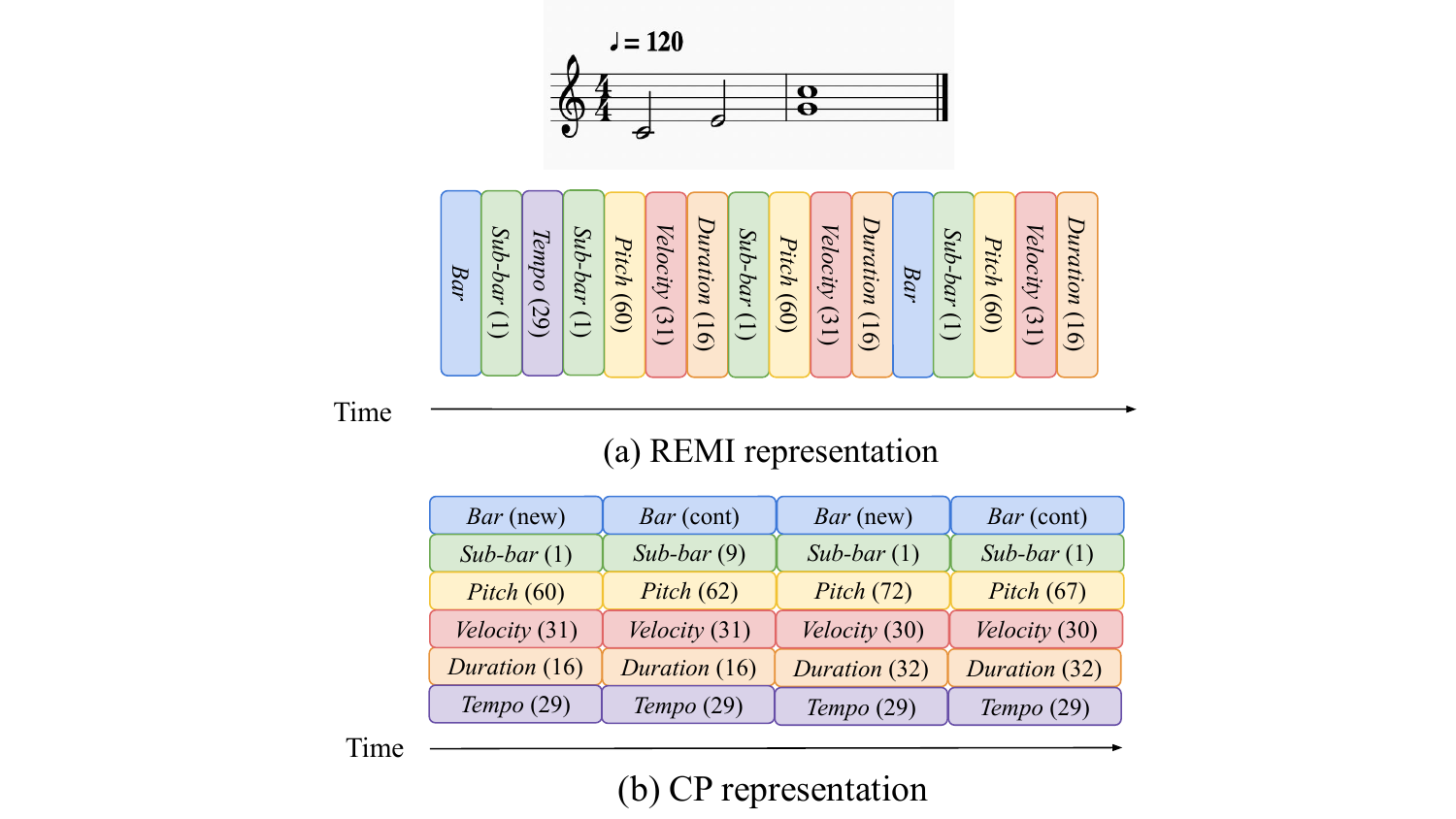}
\caption{An example of a piece of score encoded using the proposed simplified version of the (a) REMI and (b) CP representations, using seven types of tokens, \texttt{Bar}, \texttt{Sub-bar}, \texttt{Pitch}, \texttt{Velocity}, \texttt{Duration}, \texttt{Tempo} and \texttt{Pad} (not shown here), for piano-only MIDI performance.
The text inside parentheses indicates the value each token takes.
While each time step corresponds to a single token in REMI, each time step would correspond to a \emph{super token} that assembles four tokens in total in CP. Without such a token grouping, the sequence length (in terms of the number of time steps) of REMI is longer than that of CP (in this example, 16 \textit{versus} 4). Please note that the actual scores employed in our work are not as simple as this example as they are polyphonic.}
\label{fig:1}
\end{figure}

In this work, we adopt the beat-based REMI token representation proposed by \textcite{huang2020pop} to place musical notes over a discrete time grid comprising 16 equidistant sample points per bar.
In addition to REMI, we experiment with the ``token grouping'' idea of the \emph{compound word} (CP) representation \parencite{hsiao21aaai}, to reduce the length of the token sequences.
We depict the two adopted token representations in Fig. \ref{fig:1} and provide some details below.

\subsubsection{REMI Token Representation}

The REMI representation \parencite{huang2020pop} for MIDI performances uses \texttt{Bar} and \texttt{Sub-bar} tokens to represent the advancement in time. The former marks the beginning of a new bar, while the latter points to a discrete position within a bar. Specifically, as we divide a bar into 16 equidistant sample points, the \texttt{Sub-bar} tokens can take values from 1 to 16; e.g., \texttt{Sub-bar(1)} indicates the position corresponding to the first sample point in a bar, i.e., the first beat in 4/4 time signature, whereas \texttt{Sub-bar(9)} indicates the third beat.\footnote{We note that \textcite{huang2020pop} originally referred to such \texttt{Sub-bar} tokens as \texttt{Position} tokens, while \textcite{themetransformer} and \textcite{musemorphose} call them \texttt{Sub-beat} tokens. We prefer our naming for it is musically more accurate---our \texttt{Sub-bar} tokens are subdivisions of a bar (i.e., dividing a bar into 16 points), not subdivisions of a beat (i.e., not dividing a beat into 16 points).} As depicted in Fig. \ref{fig:1}(a), we use a \texttt{Sub-bar} token before each musical note, which comprises two consecutive tokens of \texttt{Pitch} and \texttt{Duration}. In other words, the \texttt{Sub-bar} token indicates the onset time of a note played at a certain MIDI pitch (i.e., the value taken by the \texttt{Pitch} token), whose duration is indicated by the \texttt{Duration} token, in multiples of demisemiquavers.
For example, \texttt{Duration(1)} and \texttt{Duration(32)} correspond to a thirty-second note and a whole note, respectively. 
For MIDI performances, a musical note is represented by not only \texttt{Pitch} and \texttt{Duration} tokens but also a \texttt{Velocity} token that indicates how hard this note was pressed by key. 
Moreover, we use the \texttt{Tempo} token to specify the tempo information of the songs. It is placed behind the \texttt{Sub-bar} token to imply when the song would perform with the tempo. We only add tempo token at the beginning of the song and the timing when tempo changes. For MIDI scores, the \texttt{Velocity} and \texttt{Tempo} tokens are simply dropped.

\subsubsection{CP Token Representation}

Fig. \ref{fig:1}(a) shows that, except for \texttt{Bar}, the other tokens in a REMI sequence always occur consecutively in groups, in the order of \texttt{Sub-bar}, \texttt{Pitch}, \texttt{Duration}. We can further differentiate \texttt{Bar(new)} and \texttt{Bar(cont)}, representing respectively the beginning of a new bar and a continuation of the current bar and always have one of them before a \texttt{Sub-bar} token. This way, the tokens would always occur in a group of four for MIDI scores. 
For MIDI performances, six tokens would be grouped together, including \texttt{Velocity} and \texttt{Tempo}.  Following the logic of \texttt{Bar}, if there is no tempo change, we simply repeat the tempo value.
Instead of feeding the token embedding of each of them individually to the Transformer, we can combine the token embedding of either the four tokens for MIDI scores or six tokens for MIDI performances in a group by concatenation and let the Transformer model
process them jointly, as depicted in Fig. \ref{fig:1}(b). We can also modify the output layer of the Transformer so that it predicts multiple tokens at once with different heads.
These constitute the main ideas of the CP representation \parencite{hsiao21aaai},  
which has at least the following two advantages over its REMI counterpart: 1) the number of time steps needed to represent a MIDI piece is much reduced, since the tokens are merged into a ``super token'' (a.k.a. a ``compound word''  \parencite{hsiao21aaai}) representing four tokens at once; 2) the self-attention in Transformer is operated over the super tokens, which might be musically more meaningful as each super token jointly represents different aspects of a musical note.  
Therefore, we experiment with both REMI and CP in our experiments.

\subsubsection{On Zero-padding}

To train Transformers, it is required that all input sequences have the same length. For both REMI and CP, we divide the token sequence for each entire piece into a number of shorter sequences with equal sequence length 512, zero-padding those at the end of a piece to 512 with an appropriate number of \texttt{Pad} tokens.
Because of the token grouping, a CP sequence for the Pop1K7 dataset would cover around 25 bars on average, whereas a corresponding REMI sequence covers only 9 bars on average.

For MIDI scores, our final token vocabulary for REMI contains 16 unique \texttt{Sub-bar} tokens, 86 \texttt{Pitch} tokens,  64 \texttt{Duration} tokens, one \texttt{Bar} token, one \texttt{Pad} token and one \texttt{Mask} token, in total 169 tokens. For CP, we do not use a \texttt{Pad} token but represent a zero-padded super token by \texttt{Bar(Pad)}, \texttt{Sub-bar(Pad)}, \texttt{Pitch(Pad)} and \texttt{Duration(Pad)}. 
We do similarly for a masked super token, using \texttt{Bar(Mask)}, etc.
Adding that we need an additional bar-related token \texttt{Bar(cont)} for CP, so the vocabulary size for CP is 169$-$2$+$8$+$1$=$176. 
For MIDI performances, the vocabulary sizes are 299 and 310 using the REMI and CP representations, respectively.

\section{Task Specification}
\label{sec:task}

Throughout this article, we refer to \emph{note-level classification tasks} as tasks that perform a prediction for each individual note in a music sequence and \emph{sequence-level tasks} as tasks that require a single prediction for an entire music sequence. We consider two note-level tasks and two sequence-level tasks in our experiments, as elaborated below.

\subsection{Symbolic-domain Melody Extraction}
\label{sec:task:melody}

For symbolic-domain melody extraction, initial methodologies predominantly adopted rule-based approaches. These rule-based methods encompassed techniques such as utilising pitch contour characteristics \parencite{melody1}, as well as the implementation of the ``skyline" algorithm \parencite{chia01skyline}. 
In recent years, deep learning-based approaches utilising convolutional neural networks (CNN) have been adopted \parencite{simonettaCNW19,note-affinity}, We will review such CNN-based methods in Section \ref{sec:baseline}, highlighting their specific details and implementation.
Similar to \textcite{simonettaCNW19}, we regard \emph{melody extraction} as a task that identifies the melody notes in a single-track~\footnote{It is common for MIDI files to consist of multiple tracks. We refer to ``single-track" as MIDI files containing only one track, which is in contrast to multi-track MIDI files that have multiple tracks.} homophonic or polyphonic music.
Utilising the POP909 dataset \textcite{pop909}, we can develop a model that classifies each Pitch event into vocal melody, instrumental melody or accompaniment, with classification accuracy (ACC) serving as the evaluation metric.\footnote{We note that there is a task closely related to melody extraction, called \emph{melody track identification}. The goal of this task is to distinguish the melody track from other non-melody tracks present in a multi-track MIDI file \parencite{madsen07IWAIM,jiang19smc}. While melody extraction is a note-level classification task, melody track identification is a track-level task. The latter is also an important symbolic music classification task, but we do not consider it here for we exclusively focus on piano-only data.}

Specifically, we consider two formulations of the task. Firstly, we adhere to the original configuration of POP909 and perform \textbf{three-class} melody classification, classifying each Pitch into three categories: vocal melody, instrumental melody or accompaniment. Secondly, we merge vocal melody and instrumental melody into a general "melody" category (while accompaniment is designated as "non-melody") and perform \textbf{two-class} classification.  Doing so allows for a direct comparison with established baselines, such as the skyline algorithm and the baseline introduced in Section \ref{sec:baseline}. For detailed results and a thorough examination, please refer to Section \ref{sec:exp:main:melody}.

\subsection{Symbolic-domain Velocity Prediction}

Dynamics is an important element in music, as they are often used by musicians to add excitement and emotion to songs. Given that the tokens we choose do not contain performance information, it is interesting to see how a machine model would ``perform'' a piece by deciding these volume changes, a task that is essential in  \emph{performance generation} \parencite{widmer94aaai, jeongKKLN19ismir, jeongKKN19icml} or \emph{expressive performance modelling} \parencite{Friberg2006b, Friberg2007}. In the realm of MIDI, velocity is a parameter that scales the intensity or volume at which a sound sample is played back, with the value ranging from 0 to 127. Default MIDI velocity values are associated with dynamic indications. Apple's Logic Pro 9 user manual correlates traditional volume indicators (\texttt{pp}, \texttt{p}, \texttt{mp}, \texttt{mf}, \texttt{f}, \texttt{ff} and \texttt{fff}) with specific MIDI velocity values (16, 32, 48, 64, 80, 96, 112 and 127), respectively.\footnote{\url{https://help.apple.com/logicpro/mac/9.1.6/en/logicpro/usermanual/} (page 468 in the user manual; accessed 2023-06-22)} In our work, we define and classify this information into six categories: \texttt{pp} (0--31), \texttt{p} (32--47), \texttt{mp} (48--63), \texttt{mf} (64--79), \texttt{f} (80--95) and \texttt{ff} (96--127). Our definition aligns with the Logic Pro 9 specifications, except that we treat \texttt{fff} as equivalent to \texttt{ff}.
Our objective can be treated as a note-level classification task, aiming to classify \texttt{Pitch} events into six classes using the POP909 dataset \parencite{pop909}.

\subsection{Symbolic-domain Style Classification}

Genre classification \parencite{correa16survey} can be considered as a type of style classification.
While genre classification categorises music based on shared musical attributes and conventions, style classification seeks to capture the nuanced stylistic variations within either a specific genre, composer or performer, accounting for the diverse artistic choices and performance practices that shape musical expressions.
We could relatively more easily find out which type of music we are listening to based on the similar patterns in that genre, while needing more musical insights to recognise the composer's or performer's style. 
Deep learning-based composer classification in MIDI has been attempted by \textcite{lee20ismirLBD} and \textcite{kong2020largescale}, both treating MIDI pieces as 2D-representation matrices (via the piano-roll representation) and using CNN classifiers. Our work differs from theirs in that: 1) we encode MIDI pieces as token sequences, 2) we employ PTM, 3) we consider non-classical music pieces and 4) our task is about style classification because not all the pianists in Pianist8 composed the pieces they performed.

\subsection{Symbolic-domain Emotion Classification}

Emotion classification in MIDI has been approached by a few researchers, mostly using hand-crafted features and non-deep learning classifiers \parencite{grekow2009detecting,lin2013exploration,panda2013multi,panda2018}. 
Some researchers work on MIDI alone, while others use both audio and MIDI in multi-modal emotion classification \parencite{panda2013multi}.
The only deep learning-based approach we are aware of is presented by \textcite{emopia}, using an RNN-based classifier called ``Bi-LSTM-Attn'' \parencite{lin2017structured} but without employing PTMs, which is also used as a baseline in our experiment; see Section \ref{sec:baseline}.

\section{Baseline Model}
\label{sec:baseline}

For the note-level classification tasks, we use an RNN model as our baseline that consists of three bi-directional long short-term memory (Bi-LSTM) layers, each with 256 neurons and a feed-forward %,dense 
layer for classification, since such a network has led to state-of-the-art result in many audio-domain music classification tasks, like beat tracking \parencite{madmom,chiu21spl} and pitch estimation \parencite{hawthorne2018onsets}.
All of our downstream tasks can be viewed as a multi-class classification problem. Given a REMI sequence, a Bi-LSTM model makes a prediction for each \texttt{Pitch} token, ignoring all the other types of tokens (i.e., \texttt{Bar}, \texttt{Sub-bar}, \texttt{Duration} and \texttt{Pad}). For CP, the Bi-LSTM model simply makes a prediction for each super token, again ignoring the zero-padded ones.

For the sequence-level tasks, which require only a prediction for an entire sequence, we follow \textcite{emopia} and choose the Bi-LSTM-Attn model from \textcite{lin2017structured} as our baseline, which was originally proposed for sentiment classification in NLP.
The model combines LSTM with a self-attention module for temporal aggregation. Specifically, it uses a Bi-LSTM layer to convert the input sequence of tokens into a sequence of embedding, which can be considered as feature representations of the tokens and then fuses these embeddings into one sequence-level embedding according to the weights assigned by the attention module to each token-level embedding. The sequence-level embedding then goes through two dense layers for classification.  
We use the token-level embedding for all the tokens here.

For melody extraction, we implement additionally the skyline algorithm \parencite{chia01skyline} and a CNN-based method \parencite{simonettaCNW19} for performance comparison.
The skyline algorithm can only perform ``melody \textit{versus} non-melody'' two-class classification for it cannot distinguish between vocal melody and instrumental melody---it uses the simple rule of taking the note with the highest pitch among the concurrent notes as the melody, while avoiding temporally overlapping notes \parencite{chia01skyline}.

The CNN method \parencite{simonettaCNW19} uses piano-roll, a 2D representation where the x-axis represents symbolic time and the y-axis represents pitch, to represent MIDI. 
Their CNN learns to predict the probability that each note belongs to the melody line. Then, a clustering algorithm is used to find a threshold for each piece adaptively. Finally, the Bellman-Ford algorithm is adopted to pick a strictly monophonic melody line.
In contrast, we do not have postprocessing steps such as thresholding or clustering in our BERT-based model and the RNN baseline.

The source code for Simonetta's model \parencite{simonettaCNW19} is available online\footnote{\url{https://github.com/LIMUNIMI/Symbolic-Melody-Identification}} but we make the following modifications to improve the model's performance:  
we use binary cross-entropy loss instead of mean error loss, sigmoid rather than ReLU activations, 
an Adam optimizer with learning rate 1e-4, 
and dropout to prevent overfitting.  
We share the re-implemented version online.\footnote{\url{https://github.com/sophia1488/symbolic-melody-identification}}

As an additional baseline for style and emotion classification, we implement the ResNet50-based CNN model from \textcite{lee20ismirLBD}, which represents the state-of-the-art for composer classification, based on the authors' code.\footnote{\url{https://github.com/KimSSung/Deep-Composer-Classification}}

\begin{figure*}
\centering
\includegraphics[trim=30 3 7 8, clip,width=\textwidth]{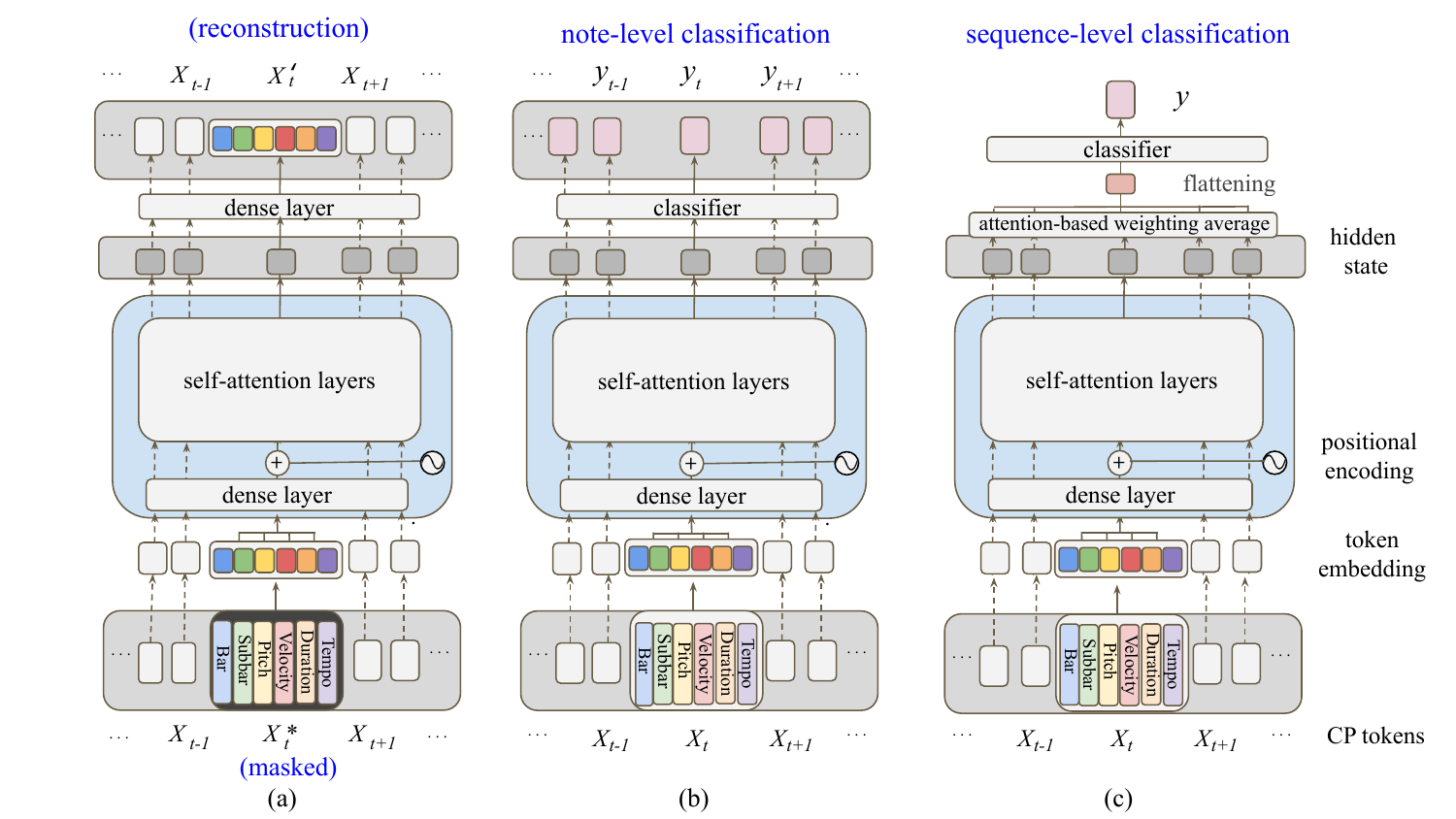}
\caption{Illustration of the (a) pre-training procedure of our model for a CP sequence, where the model learns to predict (reconstruct) randomly-picked super tokens masked in the input sequence (each consisting of four tokens, as the example one shown in the middle with time step $t$); and
(b),\,(c) the fine-tuning procedure for note-level and sequence-level classification. Apart from the last few output layers, both pre-training and fine-tuning use the same architecture.
}  
\label{fig:sys}
\end{figure*}

\section{BERT Pre-training and Fine-tuning}
\label{sec:bert}

We now present our PTM, a pre-trained Transformer encoder with 111M parameters for piano MIDI music.  
We adopt as the model backbone the BERT$_\text{BASE}$ model \parencite{bert}, a classic multi-layer bi-directional Transformer encoder with 12 layers of multi-head self-attention, each with 12 heads and the dimension of the hidden space of the self-attention layers being 768.  
Below, we first describe the pre-training strategy, then the fine-tuning method for the downstream tasks.

\subsection{Pre-training}
\label{sec:bert:pretrain}

For PTMs, an unsupervised or self-supervised, pre-training task is needed to set the objective function for learning. 
We employ the masked language modelling (MLM) pre-training strategy of BERT, randomly masking 15\% tokens of an input sequence and the Transformer will reconstruct these masked tokens from the context of the visible tokens by minimising the cross-entropy loss.
As a self-supervised method, MLM needs no labelled data relating to the downstream tasks for pre-training. Following BERT, among all the masked tokens, we replace 80\% by \texttt{MASK} tokens, 10\% by a randomly chosen token and leave the last 10\% unchanged.
Doing so has the effect of helping mitigate the mismatch between pre-training and fine-tuning as \texttt{MASK} tokens do not appear at all during fine-tuning.
For REMI, we mask the individual tokens at random. For CP, we mask the super tokens---when we mask a super token, we have to reconstruct all the four tokens composing it by different output heads \parencite{hsiao21aaai}, as shown in Fig. \ref{fig:sys}(a).

There are three steps for processing the input token. First, each input token $X_i$ is converted into a token embedding $E_i$ through an embedding layer. Second, it is augmented by addition with a relative positional encoding \parencite{relative-key-query} that is related to its time step. Third, it is then fed $E_i$ to the stack of 12 self-attention layers to get a ``contextualised'' representation known as a \emph{hidden vector} at the output of the self-attention stack.
Because of the bi-directional self-attention layers, the hidden vector is contextualized in the sense that it has attended to information from all the other tokens from the same sequence. 
Finally, the hidden vector of a masked token
is fed into a dense layer to predict what the missing super token is.
As our network structure is rather standard, we refer readers to the literature \parencite{vaswani2017attention,bert,musemorphose} for details and the mathematical underpinnings due to space limits. Because the vocabulary sizes for the four token types are different, we weight the training loss associated with tokens of different types in proportion to the corresponding vocabulary size of REMI and CP, to facilitate model training. 

We note that the original BERT article also used another self-supervised task called ``next sentence prediction'' (NSP) \parencite{bert} together with MLM for pre-training.  We do not use NSP for our model since it was later shown to be not that useful \parencite{spanbert,xlnet,cross-lingual-lm}; moreover, NSP requires a clear definition of ``sentences'', which is not well-defined for our MIDI sequences. As a result, we do not use tokens such as \texttt{CLS}, \texttt{SEP} and \texttt{EOS} used by BERT for making the boundary of the sentences.\footnote{
\texttt{CLS} marks the beginning of a sentence, \texttt{SEP} the boundary between two consecutive sentences (useful for the so-called ``next sentence prediction task'' \parencite{bert}) and \texttt{EOS} the end of a sentence.}

\subsection{Fine-tuning}
\label{sec:bert:finetune}

It has been widely shown in NLP and related fields \parencite{speechbert,vilbert,videobert,proteinbert} that, by storing knowledge in huge numbers of parameters and carrying out task-specific fine-tuning,
the knowledge implicitly encoded in the parameters of a PTM can be transferred to benefit a variety of downstream tasks \parencite{han2021pretrained}. 
For fine-tuning, we extend the architecture shown in Fig. \ref{fig:sys}(a) by modifying the last few layers in two different ways, one for each of the two types of downstream tasks.

Fig. \ref{fig:sys}(b) shows the fine-tuning architecture for note-level classification. While the Transformer uses the hidden vectors to recover the masked tokens during pre-training, it has to predict the label of an input token during fine-tuning, by learning from the labels provided in the training data of the downstream task in a supervised way.
To achieve this, we feed the hidden vectors to a stack of dense layers, a ReLU activation layer and finally another dense layer for the output classification, with 10\% dropout probability. 
We note that this classifier design is fairly simple, 
as we expect much knowledge regarding the downstream task can already be extracted from the preceding self-attention layers.

Fig. \ref{fig:sys}(c) shows the fine-tuning architecture for sequence-level classification. 
Being inspired by the Bi-LSTM-Attn model  \parencite{lin2017structured}, we employ an attention-based weighting average mechanism to convert the sequence of 512 hidden vectors for an input sequence to one single vector before feeding it to the classifier layer, which comprises two dense layers. 
We note that, unlike the baseline models introduced in Section \ref{sec:baseline}, we do not use RNN layers in our models. An alternative approach is to add the \texttt{CLS} token to our sequences and simply use its hidden vector as the input to the classifier layer. We do not explore this alternative since we do not have \texttt{CLS} tokens.

\subsection{Implementation Details}

Our implementation is based on the open-source library HuggingFace \parencite{huggingface}.
As mentioned in Section \ref{sec:db2}, we use Pop1K7 and ASAP for pre-training and the other three datasets (i.e., POP909, Pianist8 and EMOPIA) for the downstream tasks.
From the combination of Pop1K7 and ASAP, we use 85\% of them for pre-training as described in Section \ref{sec:bert:pretrain} and the rest as the validation set. 
We train with a batch size of 12 sequences for at most 500 epochs (i.e., around 500K iterations for REMI and 1M iterations for CP), using the AdamW optimizer with learning rate 2e$-$5 and weight decay rate 0.01. 
If the validation cross-entropy loss does not improve for 30 consecutive epochs, we stop the training process early. For pre-training, we can improve the validation ``cloze'' accuracy from 70.4\% for REMI to 78.73\% for CP.
We observe that pre-training using the CP representation converges in 2.5 days on four GeForce GTX 1080-Ti GPUs, which is about 2.5 times faster than the case of REMI.
Moreover, a smaller batch size degrades overall performance, including downstream classification accuracy.
Because each sequence has 512 super tokens, we have 6,144 super tokens per batch.

For fine-tuning, we create training, validation and test splits for each of the three datasets of the downstream tasks with the 8:1:1 ratio at the piece level (i.e., all the 512-token sequences from the same piece are in the same split). 
With the same batch size of 12, we fine-tune the pre-trained our model for each task independently for at most 10 epochs, early stopping when there is no improvement for three consecutive epochs.  Compared to pre-training, fine-tuning is less computationally expensive.  All the results reported in our work can be reproduced with four GeForce GTX 1080-Ti GPUs within 30 minutes.

In our experiments, we will use the same pre-trained model parameters to initialise the models for different downstream tasks. During fine-tuning, we fine-tune the parameters of all the layers, including the self-attention and token embedding layers. 

\begin{table}[t]
\caption{The testing classification accuracy (in \%) of different combinations of MIDI token representations and models for  four downstream tasks: three-class melody classification, velocity prediction, style classification and emotion classification. ``CNN'' represents the ResNet50 model used by \textcite{lee20ismirLBD}, which only supports sequence-level tasks. ``RNN'' denotes the baseline models introduced in Section \ref{sec:baseline}, representing the Bi-LSTM model for the first two (note-level) tasks and the Bi-LSTM-Attn model \parencite{lin2017structured} for the last two (sequence-level) tasks. } 
\label{tab:exp}
\begin{center}
\begin{tabular}{l l c c c c}
\toprule
Token~~~~~~~~~~~~ & Model & Melody~~ & Velocity~~ & Style~~ & Emotion \\
\midrule
\multirow{4}{2em}{REMI} & CNN \parencite{lee20ismirLBD} & --- & --- & 51.35 & 60.00 \\
& RNN \parencite{lin2017structured}~ & 89.96 & 44.56 & 51.97 & 53.46 \\
& Our model\,(\texttt{score}) & 90.97 & 49.02 & 67.19 & 67.74\\
& Our model\,(\texttt{performance}) & 89.23 & --- & 75.30 & 66.18\\
\midrule
\multirow{4}{2em}{CP} & CNN \parencite{lee20ismirLBD} & --- & --- & 67.57 & 60.00 \\
& RNN \parencite{lin2017structured} & 88.66 & 43.77 & 60.32 & 54.13 \\
& Our model\,(\texttt{score}) & \textbf{96.15} & \textbf{52.11} & 67.46 & 64.22\\
& Our model\,(\texttt{performance}) & 95.83 & --- & \textbf{81.75} & 70.64 \\
\midrule
\multirow{1}{2em}{OctupleMIDI} & MusicBERT \parencite{musicbert} & --- & --- & 37.25 & \textbf{77.78} \\
\bottomrule
\end{tabular}
\end{center}
\end{table}

\section{Performance Study}
\label{sec:exp}

In what follows, we use `our model\,(\texttt{score})' to indicate the result when MIDI scores are considered and similarly  `our model\,(\texttt{performance})' for MIDI performances. 
Since MIDI performance contains \texttt{velocity} information, we do not evaluate on the velocity prediction task for fairness. We note that, while `our model\,(\texttt{score})' and `our model\,(\texttt{performance})' adopt different token representations, we consider it valid to compare their performance as their training and test data are respectively from the same sets of music pieces.

Tab. \ref{tab:exp} lists the testing accuracy achieved by the baseline models and the proposed ones for  four downstream tasks.
We see that ``our model\,(\texttt{score})'' outperforms the Bi-LSTM or Bi-LSTM-Attn baselines in all tasks consistently, using either the REMI or CP representation.  
In particular, the combination of our model (score) and CP, referred to as ``our model (score)+CP'' hereafter, exhibits the highest accuracy in the two note-level tasks. Additionally, the combination of our model (performance) and CP, denoted as ``our model (performance)+CP'', achieves the best result in the style classification task, while demonstrating a notable improvement in accuracy compared to REMI for emotion classification.
We also observe that our models outperform Bi-LSTM$+$CP with just 1 or 2 epochs of fine-tuning, validating the strength of PTMs on symbolic-domain music classification tasks.

To facilitate a comprehensive evaluation, we additionally incorporate a officially released version of MusicBERT \parencite{musicbert} in the sequence-classification tasks. 
Specifically, we use the model checkpoint MusicBERT-small,\footnote{\url{https://github.com/microsoft/muzic/tree/main/musicbert}} which is pre-trained on the Lakh MIDI (LMD) dataset \parencite{MIDIfesto}, which contains about 100K songs.\footnote{There is another implementation named MusicBERT-base, which is trained on the Million MIDI Dataset \parencite{musicbert}, which is ten times larger than LMD.} 
The results show that MusicBERT achieves a testing accuracy of 37.25\% for style classification and 77.78\% for emotion classification. Specifically, in the style classification task, MusicBERT exhibits clear signs of overfitting and falls short in performance when compared to our model (81.75\%). This outcome can be attributed to the limited size of the Pianist8 dataset, comprising only 411 songs. Conversely, in the emotion classification task, MusicBERT demonstrates impressive performance, surpassing our model (70.64\%) by a significant margin. This finding is intriguing and suggests that the application of large-scale pre-training may yield substantial benefits in  classifying the emotional content of a MIDI piece.

Tab. \ref{tab:exp} also shows that the CP token representation tends to outperform the REMI one across different tasks for both the baseline models and the PTM-based models, demonstrating the importance of token representation for music applications.  
To study whether the accuracy gain comes simply from a longer musical context enjoyed by CP, we also train ``our model\,(\texttt{performance})$+$CP'' with a sequence of length 128, obtaining 95.43, 80.32 and 64.04 accuracies for three-class melody classification, style classification and emotion classification, respectively.  
We note a sequence of length 512 for REMI 
contains approximately the same amount of information for a sequence of 147 supertokens for CP.  Still, using the CP token representation in general leads to better performance even with less information.

Tab. \ref{tab:exp} also shows that ``our model\,(\texttt{performance})$+$CP'' outperforms ``our model\ (\texttt{score})$+$CP'' greatly for the two sequence-level tasks, style classification and emotion classification. This matches our intuition as the two tasks are highly related to performance styles and expressions of the piano pieces.

We take a closer look at the performance of the evaluated models, in particular Bi-LSTM$+$CP (or Bi-LSTM-Attn$+$CP), ``our model\,(\texttt{score})$+$CP'' and ``our model\,(\texttt{performance})$+$CP'' in different tasks in what follows.

\begin{figure}[t]
\centering
\includegraphics[trim=10 45 9
95, clip,width=\textwidth]{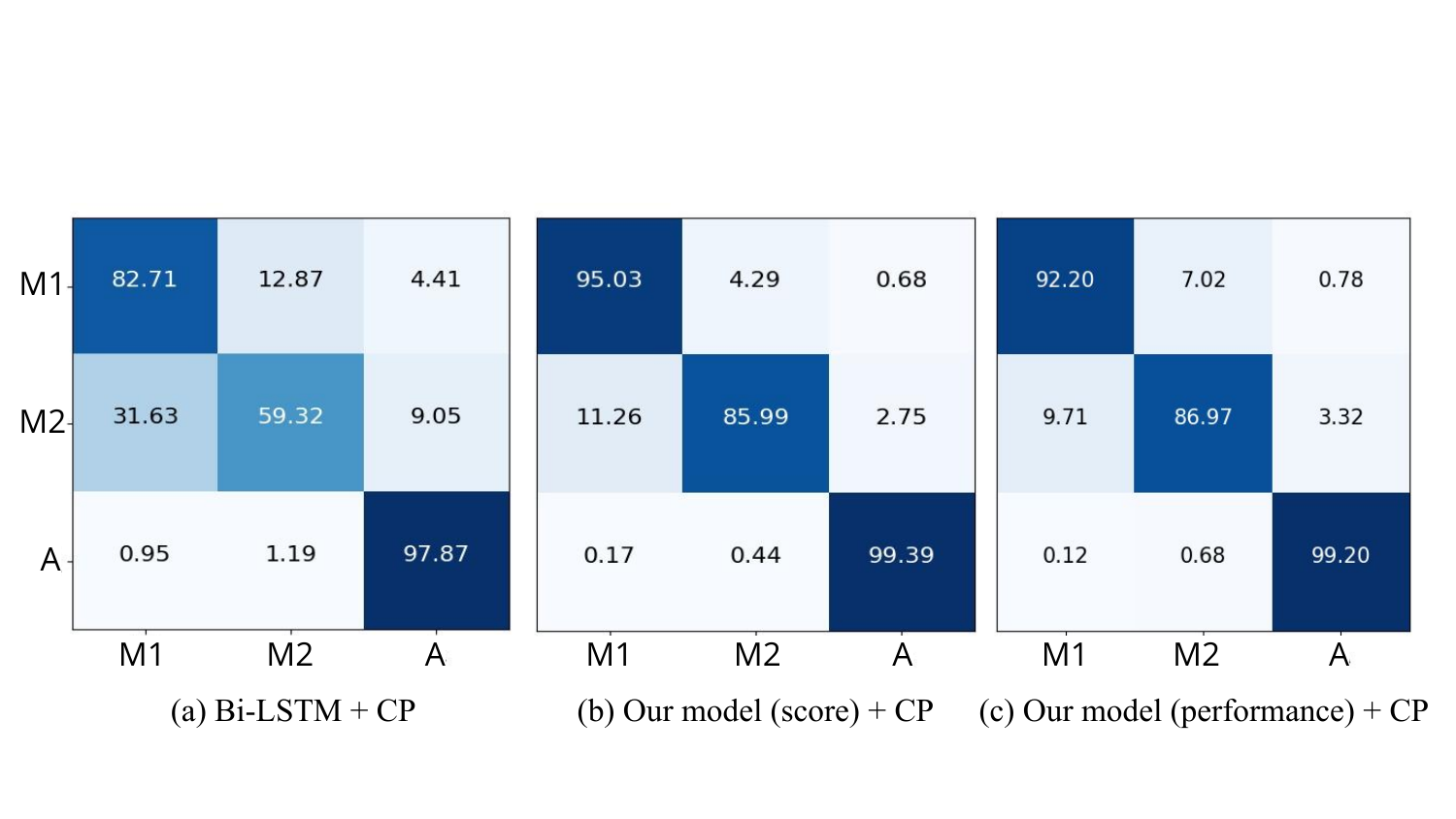}
\caption{Confusion tables (in \%) for two models for three-class melody classification, calculated on the test split of POP909$_\text{4/4}$. Each row represents the percentage of notes in an actual class while each column represents a predicted class. Notation---``M1'': \texttt{vocal melody}, ``M2'': \texttt{instrumental melody}, ``A'': \texttt{accompaniment}.}
\label{fig:confusion-melody}
\end{figure}

\begin{table}[t]
\caption{Testing metrics (in \%) of ``our model (performance) +CP'' and other baseline methods for the two-class ``melody \textit{versus} non-melody'' classification task using POP909, viewing  \texttt{vocal melody} and \texttt{instrumental melody} as ``melody'' and \texttt{accompaniment} as ``non-melody''.}
\label{tab:melody}
\begin{center}
\begin{tabular}{l c c c c}
\toprule
Model & Accuracy~ & Precision~ & Recall~~ & F1  \\
\midrule
Skyline \parencite{chia01skyline} & 79.52 & 81.42 & 56.57 & 66.76 \\
Simonetta \textit{et al}.'s CNN \parencite{simonettaCNW19} & 92.08 & 88.95 & 89.30 & 89.13 \\
Our model (performance) + CP & \textbf{99.06} & \textbf{98.68} & \textbf{98.72} & \textbf{98.70} \\
\bottomrule
\end{tabular}
\end{center}
\end{table}

\begin{figure*}[t]
\centering
\includegraphics[width=\textwidth]{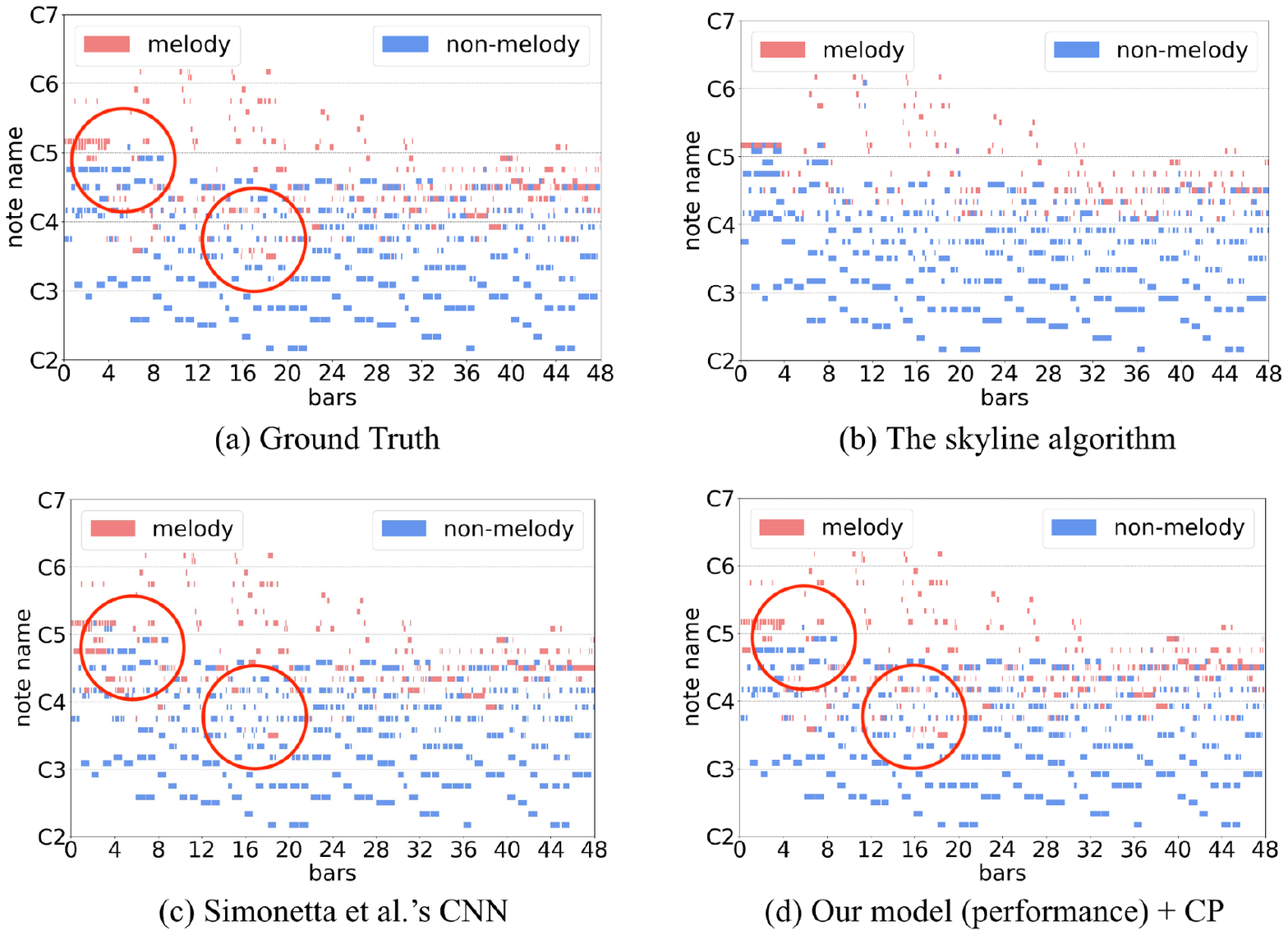}
\caption{The melody/non-melody classification result for ``POP909-596.mid'' by (b) ``skyline'' \parencite{chia01skyline}, (c) Simonetta \textit{et al}.'s CNN \parencite{simonettaCNW19} and (d) our model (performance) + CP. Directing attention to the red circled region within the pianoroll representation, it is evident that the CNN baseline faces challenges in effectively distinguishing between melody and accompaniment, particularly when note pitches reside within the C4 to C5 range during the initial phase. This is especially pronounced in low-pitch scenarios, where the CNN baseline struggles with accurate classification. In contrast, our model exhibits a notably improved predictive accuracy, closely aligning with the ground truth representation. To further supplement the information
%give a more complete picture
, the generated melody audio files and pianoroll figures are available in our repository.}
\label{fig:pr}
\end{figure*}

% \foornote{\url{https://github.com/wazenmai/MIDI-BERT/tree/CP/melody_extraction/audio}} and picture\footnote{\url{https://github.com/wazenmai/MIDI-BERT/blob/CP/melody_extraction/pianoroll/596_pianoroll.png}} is available in our repository.}

\subsection{Melody}
\label{sec:exp:main:melody}

Fig. \ref{fig:confusion-melody} presents the normalised confusion tables for three-class melody classification, illustrating distinct performance characteristics among the models. We note that the baseline exhibits a tendency to conflate vocal melody (M1) and instrumental melody (M2), whereas our model outperforms the RNN-based model,  enhancing the overall accuracy by almost 8\% (from 88.66\% to 96.15\%).
A closer examination reveals our model's superior ability to differentiate between vocal and instrumental melodies compared to the RNN baseline with minimal finetuning. This task is particularly challenging given the nature of the POP909 dataset, which exclusively features pop songs sung by humans. Consequently, the separation of vocal and instrumental melodies relies on the criterion of human vocalisation (which is absent in MIDI data), potentially leading to instances where notes between phrases are designated as instrumental melody despite sharing pitch and melodic characteristics with vocal melody. 

Interestingly, an intriguing observation emerges as ``our model\,(score)\,+\,CP'' demonstrates a more effective distinction between vocal and instrumental melodies than ``our model\,(performance)\,+\,CP''. This phenomenon suggests that even without velocity information, our model can discern segments designated for singing \textit{versus} those serving as preludes, interludes or fills.

Tab. \ref{tab:melody} compares our model with the ``skyline'' algorithm \parencite{chia01skyline} and the 
CNN-based baselines \parencite{simonettaCNW19} for the two-class ``melody \textit{versus} non-melody''  melody classification task.
As the dataset is highly unbalanced (i.e., the melody notes are  much fewer than the accompaniment notes), we also report the precision, recall and F1 scores.  It turns out that our model greatly outperforms other baselines across all the metrics, reaching 99.04\% classification accuracy.
A qualitative example demonstrating the superiority of the proposed model over the the skyline algorithm  can be found in Fig. \ref{fig:pr}, using a randomly chosen testing piece from POP909$_\text{4/4}$. 

Moreover, we have extended the application of our melody extraction model to compositions from the Pianist8 dataset. Given the absence of manual labels for the melody notes within this dataset, we encourage readers to evaluate the results by listening to the prediction outputs.\footnote{\url{https://github.com/wazenmai/MIDI-BERT/tree/CP/melody_extraction/audio}} 
We provide three versions of the melody MIDI file for each original song, generated respectively by the skyline algorithm, Simonetta \textit{et al}.'s CNN and ``our model (performance) + CP''.
Taking ``Clayderman\_Yesterday\_Once\_More.mid'' as an example, the melody generated by the skyline algorithm exhibits stiffness and lacks intricate details, retaining only the treble. The CNN version demonstrates considerable improvement over the skyline algorithm. However, a noticeable intermittent quality persists throughout the entire song, with some cohesive melodies omitted. Our model achieves commendable performance, successfully extracting the majority of the main melody and presenting a discernible melodic progression. It is worth highlighting the efficiency of our model, as it requires less than one hour for finetuning under the same hardware conditions that necessitate a full day of training for the CNN baseline on the POP909 training set.

\begin{figure}[t]
\begin{center}
\includegraphics[trim=105 33 60
65, clip,width=0.65\textwidth]{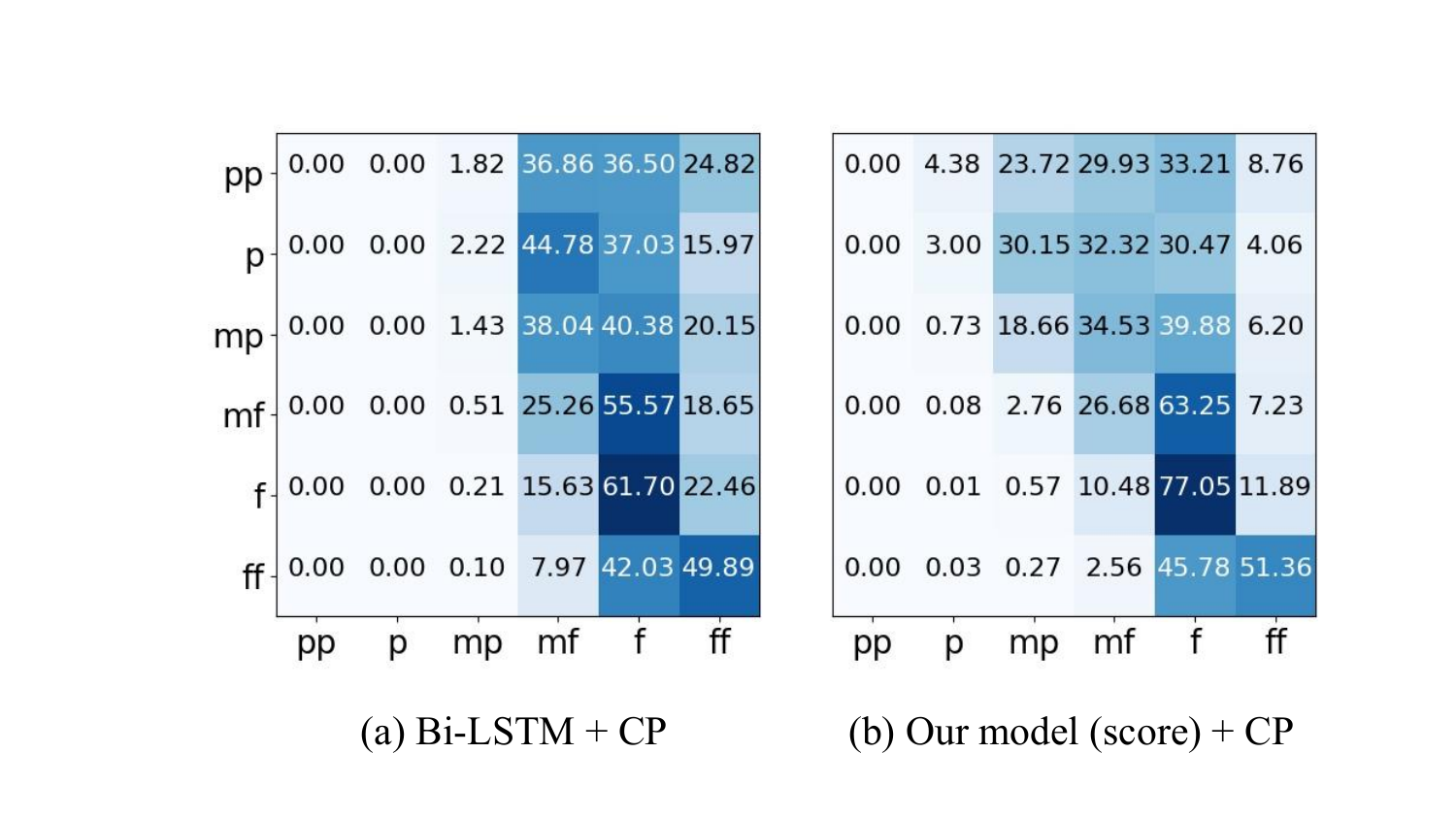}
\end{center}
\caption{Confusion tables (in \%) for velocity prediction, calculated on the test split of POP909$_\text{4/4}$. Each row represents the percentage of notes in an actual class while each column represents a predicted class.}
\label{fig:confusion-velocity}
\end{figure}

\subsection{Velocity}
\label{sec:exp:main:velocity}

Tab. \ref{tab:exp} shows that the accuracy on our 6-class velocity classification task is not high, reaching 52.11\% at best. This may be due to the fact that velocity is rather subjective, meaning that musicians can perform the same music piece fairly differently.  Moreover, we note that the data is highly imbalanced, with the latter three classes (\texttt{mf}, \texttt{f}, \texttt{ff}) taking up nearly 90\% of all labelled data.  The confusion tables presented in Fig. \ref{fig:confusion-velocity} show that Bi-LSTM tends to classify most of the notes into \texttt{f}, the most popular class among the six. 
This is less the case for our model, but the prediction of  \texttt{p} and \texttt{pp}, i.e., the two with the lowest dynamics, remains challenging. For future work, data augmentation is a potential solution to mitigate the impact of data imbalance.

\begin{figure}[t]
\begin{center}
\includegraphics[trim=4 70 20
70, clip, width=\textwidth]{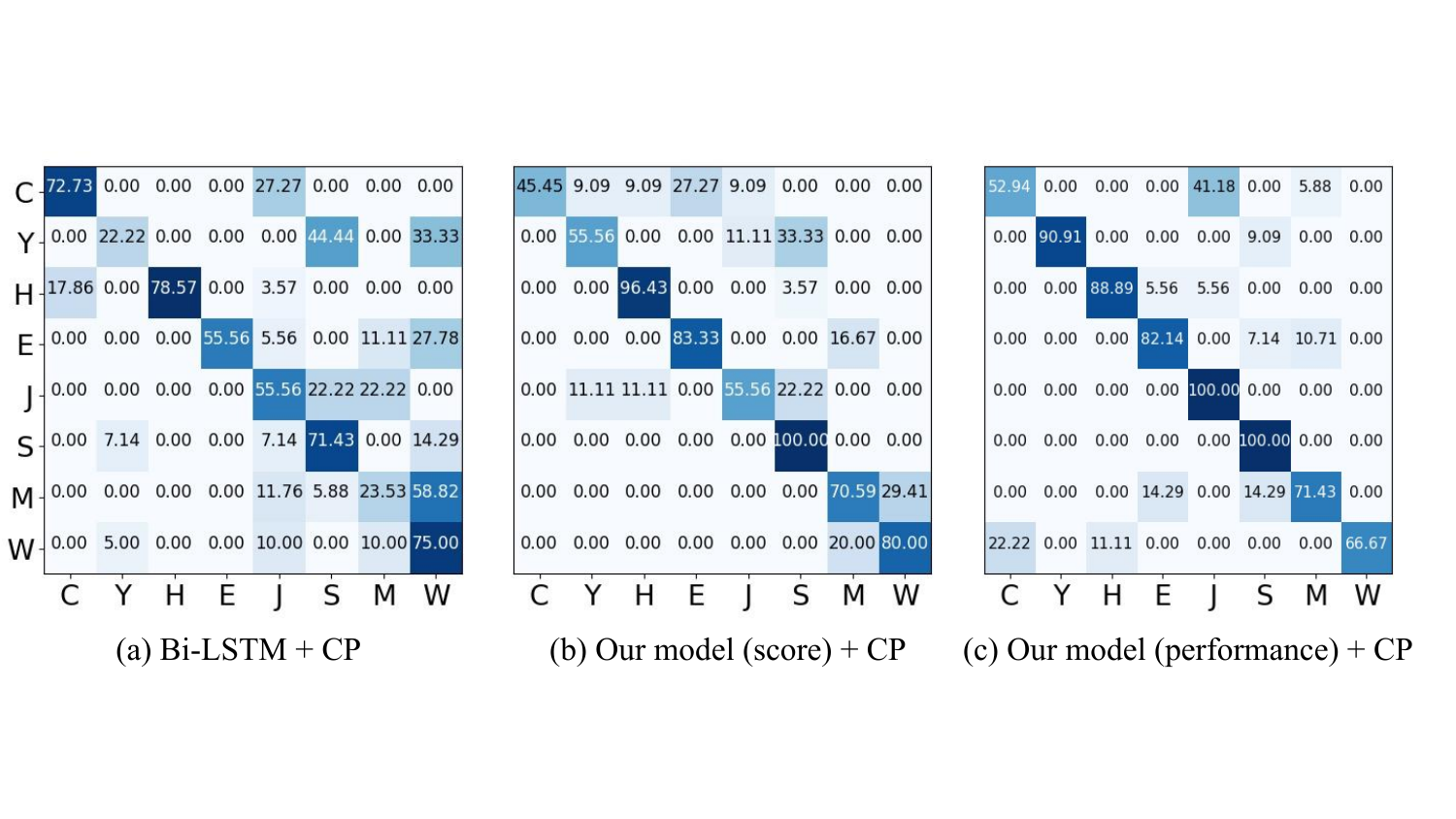}
\end{center}
\caption{Confusion tables (in \%) for style classification on the test split of Pianist8.
Each row shows the percentage of sequences of a class predicted as another class. Notation---``C'': R. Clayderman (pop), ``Y'': Yiruma (pop), ``H'': H. Hancock (jazz), ``E'': L. Einaudi (contemporary), ``J'': H. Joe (contemporary), ``S'': R. Sakamoto (contemporary), ``M'': Bethel Music (religious) and ``W'': Hillsong Worship (religious).}
\label{fig:confusion-composer}
\end{figure}

\subsection{Style}
\label{sec:exp:main:composer}

Tab. \ref{tab:exp} shows that our model greatly outperforms Bi-LSTM-Attn \parencite{lin2017structured} and the CNN baseline \parencite{lee20ismirLBD} by 10--20\% regardless of the token representation taken, reaching 81.75\% testing accuracy at best for this 8-class classification problem.
In addition, we see a large performance gap between REMI and CP in this task, the largest among the four tasks.
Fig. \ref{fig:confusion-composer} further shows that,
both the baseline and ``our model\,(\texttt{score})$+$CP'' confuse artists in similar genres and that our model performs fairly well in recognising Herbie Hancock and Ryuichi Sakamoto. In contrast, by considering velocity and tempo information, ``our model\,(\texttt{performance})$+$CP'' gains lots of precision on classifying songs in pop  and contemporary genres, boosting the classification accuracy from 67.46 (score) to 81.75 (performance).

\subsection{Emotion}
\label{sec:exp:main:emotion}

\begin{figure}[t]
\begin{center}
\includegraphics[trim=8 70 18
101, clip, width=\textwidth]{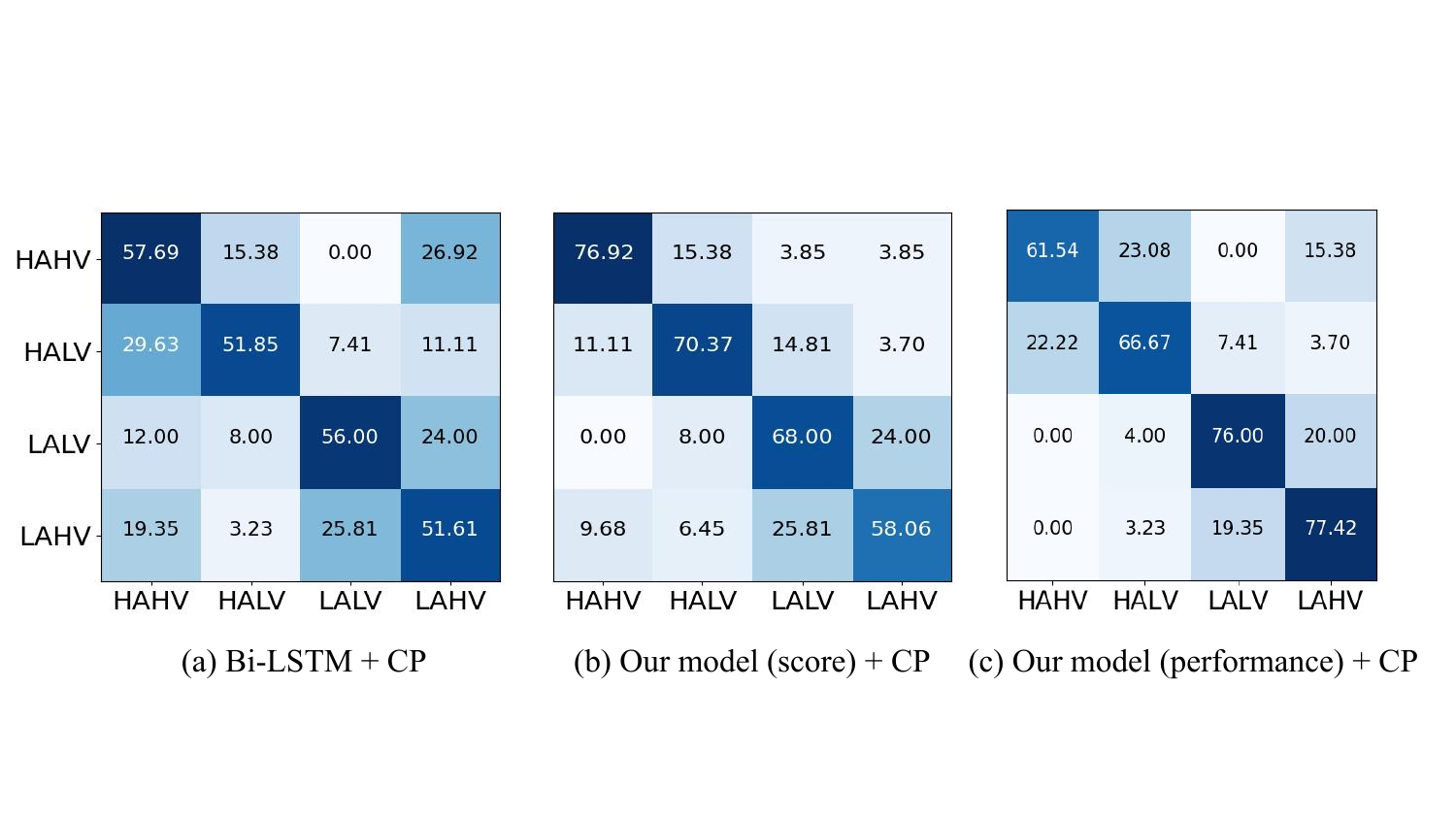}
\end{center}
\caption{Confusion tables for emotion classification; in \% of sequences on the test split of EMOPIA. Each row represents the percentage of notes in an actual class while each column represents a predicted class.}
\label{fig:confusion-emotion}
\end{figure}

Tab. \ref{tab:exp} shows that our model outscores Bi-LSTM-Attn by around 14\% and the CNN baseline \parencite{lee20ismirLBD} by around 7\% in both REMI and CP for this 4-class classification problem, reaching 70.64\% testing accuracy at best.
There is little performance difference between REMI and CP in this task. 
Fig. \ref{fig:confusion-emotion} further shows that the evaluated models can fairly easily distinguish between high arousal and low arousal pieces (i.e., ``HAHV, HALV'' \textit{versus} ``LALV, LAHV''), but they have a much harder time along the valence axis (e.g., ``HAHV'' \textit{versus} ``HALV'' and  ``LALV'' \textit{versus} ``LAHV''). We see less confusion from the result of `our model\,(\texttt{score})$+$CP'.
By considering velocity and tempo, 
``our model\,(\texttt{performance})$+$CP'' can further classify variance difference in low-arousal songs, though there is still room for improvement.

\section{Conclusion}
\label{sec:conclusion}

In this article, we presented a large-scale pre-trained model for musical data in the MIDI format. We employed five public-domain piano MIDI datasets for BERT-like masking-based pre-training and evaluated the pre-trained model on four challenging downstream symbolic music classification tasks, most with less than 1K labelled MIDI pieces. Our experiments validate the effectiveness of pre-training for both note-level and sequence-level classification tasks. 

This work can be extended in many ways. First, to employ other pre-training strategies or architectures \parencite{han2021pretrained}. 
Second, to employ Transformer models with linear computational complexity \parencite{choromanskiRethinkingAttentionPerformers2020a,liutkus21icml}, so as to use the whole MIDI pieces (instead of segments) at pre-training.\footnote{We note that the use of linear Transformers for symbolic music generation has been attempted before \textcite{hsiao21aaai}.}
Third, to include other time signatures and increase the amount of non-pop piano scores.
Fourth, to extend the corpus and the token representation from single-track piano to multi-track MIDI, like the work done by \textcite{musicbert}. Finally, to consider more downstream tasks such as symbolic-domain music segmentation \parencite{gttm14ismir,peter20ismir}, chord recognition \parencite{harasim20ismir}, score passage matching \parencite{cmerata} and beat tracking \parencite{chuang20apsipa}. We have released the code publicly, which may hopefully help facilitate such endeavours. 

% \section*{References}

% \begin{verbatim}
\printbibliography
% \end{verbatim}

\end{document}